\documentclass[11pt,a4paper,fleqn]{article}

\usepackage{amsmath,amssymb,graphicx,siunitx,multirow,float,calc,array,tabularx,caption,float}
\usepackage[table]{xcolor}
\usepackage[colorlinks=true,breaklinks=true,allcolors=blue]{hyperref}
\newcommand{\revision}[1]{#1}
\usepackage[numbers,sort&compress]{natbib}
\usepackage{authblk}

\usepackage[margin=2cm]{geometry}

\makeatletter
\def\ttabular{%
\hbox\bgroup
\let\\\cr
\def\rulea{\ifnum\rowc=\@ne \hrule height 1.3pt \fi}
\def\ruleb{
\ifnum\rowc=1\hrule height 1.3pt \else
\ifnum\rowc=6\hrule height \heavyrulewidth 
   \else \hrule height \lightrulewidth\fi\fi}
\valign\bgroup
\global\rowc\@ne
\rulea
\hbox to 10em{\strut \hfill##\hfill}%
\ruleb
&&%
\global\advance\rowc\@ne
\hbox to 10em{\strut\hfill##\hfill}%
\ruleb
\cr}
\def\endttabular{%
\crcr\egroup\egroup}

\def\figw{0.47\textwidth}
\allowdisplaybreaks[4]

\def\figw{0.18\textwidth}

\begin{document}
\title{Exponential and Weibull models for spherical and spherical-shell diffusion-controlled release systems with semi-absorbing boundaries}
\author{Elliot J. Carr\thanks{\href{mailto:elliot.carr@qut.edu.au}{elliot.carr@qut.edu.au}}}
\affil{School of Mathematical Sciences, Queensland University of Technology, Brisbane, Australia.}

\date{}

\maketitle

\section*{Abstract} 
\noindent We consider the classical problem of particle diffusion in $d$-dimensional radially-symmetric systems with absorbing boundaries. A key quantity to characterise such diffusive transport is the evolution of the proportion of particles remaining in the system over time, which we denote by $\mathcal{P}(t)$. Rather than work with analytical expressions for $\mathcal{P}(t)$ obtained from solution of the corresponding continuum model, which when available take the form of an infinite series of exponential terms, single-term low-parameter models are commonly proposed to approximate $\mathcal{P}(t)$ to ease the process of fitting, characterising and interpreting experimental release data. Previous models of this form have mainly been developed for circular and spherical systems with an absorbing boundary. In this work, we consider circular, spherical, annular and spherical-shell systems with absorbing, reflecting and/or semi-absorbing boundaries. By proposing a moment matching approach, we develop several simple one and two parameter exponential and Weibull models for $\mathcal{P}(t)$, each involving parameters that depend explicitly on the system dimension, diffusivity, geometry and boundary conditions. The developed models, despite their simplicity, agree very well with values of $\mathcal{P}(t)$ obtained from stochastic model simulations and continuum model solutions.

\section{Introduction} 
\begin{figure*}[p]
\centering
\fbox{\includegraphics[width=0.98\textwidth]{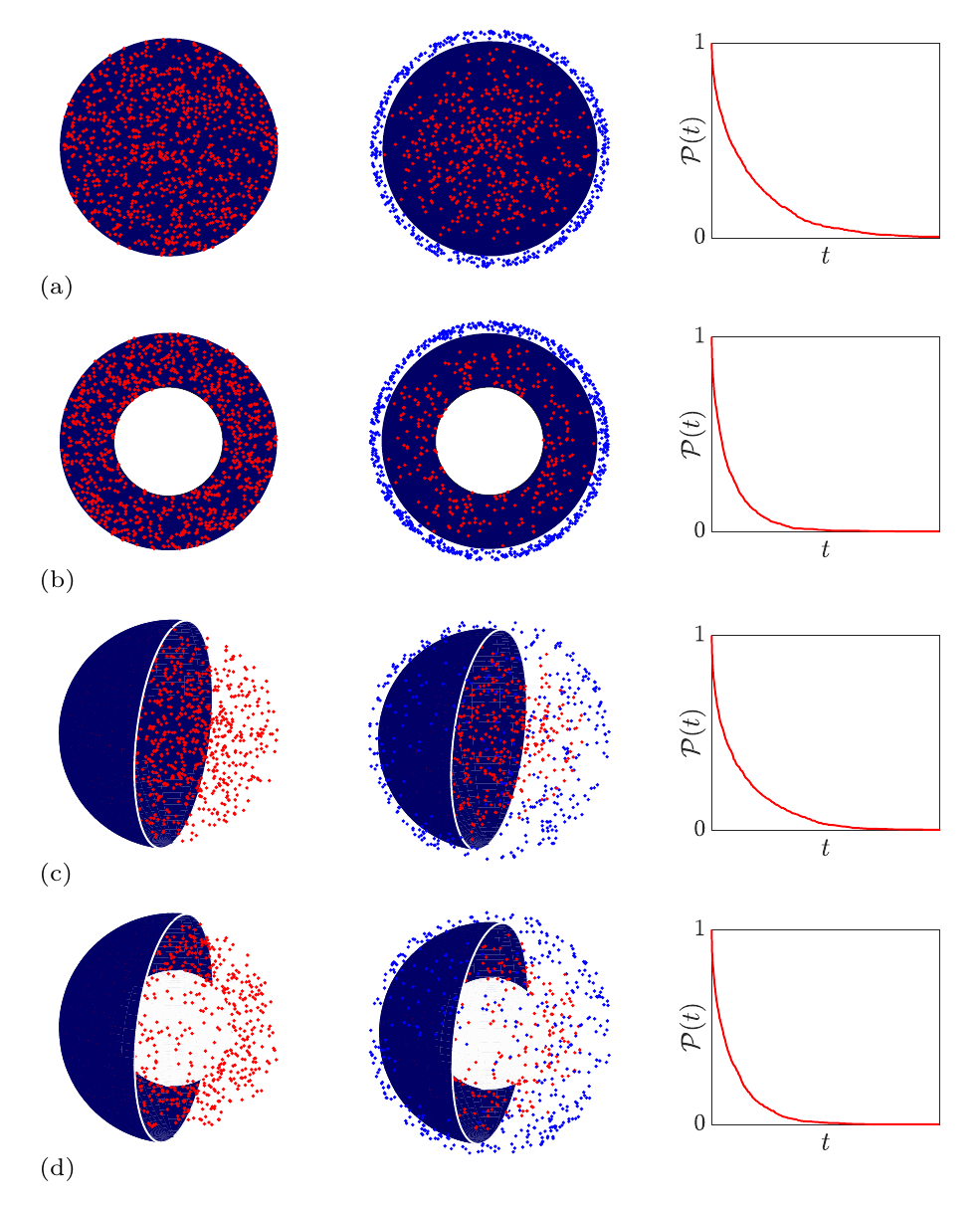}}
\caption{Diffusion controlled release from (a) a circular system with absorbing outer boundary (b) an annular system with reflecting inner boundary and absorbing outer boundary (c) a spherical system with absorbing outer boundary and (d) a spherical-shell system with reflecting inner boundary and absorbing outer boundary. Released particles are shown in blue while particles still actively diffusing in the system are shown in red. Particle distributions are shown initially and at a short time after when some of the particles have been released. For each case, the proportion of particles remaining, $\mathcal{P}(t)$, decreases over time as more particles are released from the system.}
\label{fig:figure1}
\end{figure*}

Mathematically modelling stochastic diffusive transport is fundamental to numerous applications across physics \cite{vaccario_2015,van-kampen_2007}, biology \cite{codling_2008,lotstedt_2015}, ecology \cite{okubo_2001,kurella_2015} and medicine \cite{ignacio_2021,hadjitheodorou_2013}. In this work, we consider the classical problem of particle diffusion in $d$-dimensional radially-symmetric homogeneous domains containing absorbing boundaries. Here, particles diffuse until they are absorbed at a boundary, at which point they are removed from the system (Figure \ref{fig:figure1}). A key quantity to characterise such diffusive processes is the evolution of the proportion of particles remaining in the system over time, which we denote by $\mathcal{P}(t)$ (Figure \ref{fig:figure1}). As shown in Figure \ref{fig:figure1}, $\mathcal{P}(t)$, \revision{which is called the survival probability in the first passage times literature \cite{redner_2001}}, decreases over time as more particles are released with \revision{both the slope} and the shape of the curve depending on the system dimension, geometry, diffusivity and boundary conditions.

When available, analytical expressions for $\mathcal{P}(t)$, obtained from the continuum analogue of the stochastic diffusion process, \revision{take the form of an infinite series of exponential functions \cite{simpson_2015a,ignacio_2021,carslaw_1959,crank_1975,redner_2001}} that depend non-trivially (and in some cases implicitly) on various parameters of the diffusive transport system. To ease the process of fitting and explaining experimental release data \cite{ignacio_2017}, simple one-term models are commonly proposed to approximate $\mathcal{P}(t)$. Previous work in this area has proposed exponential, Weibull and other exponential-like functions to describe $\mathcal{P}(t)$ (and related quantities in other applications) for slab, circular and spherical systems with radial symmetry and an absorbing boundary \cite{ignacio_2017,kalosakas_2015,hadjitheodorou_2013,ignacio_2021,siepmann_2008,ritger_1987,kosmidis_2003,lotstedt_2015}. In this work, we consider circular and spherical systems with an absorbing or semi-absorbing outer boundary and annular and spherical-shell systems with absorbing, reflecting or semi-absorbing inner and outer boundaries. 

\revision{Semi-absorbing boundaries \cite{collins_1949,sano_1979,sapoval_1994,grebenkov_2003} have} wide-ranging applications, e.g., drug delivery using microcapsules encased in a thin semi-permeable coating \cite{carr_2018e} and virus infection of a cell surrounded by a protective membrane \cite{chou_2007}. In this study, such boundaries are distinguished from absorbing and reflecting boundaries in that a diffusing particle passing through a semi-absorbing boundary is absorbed and removed from the system with a certain specified probability otherwise it is reflected \cite{grebenkov_2003,erban_2007}. \revision{This treatment assumes a rigid membrane/coating with fixed porosity. An alternative approach that we do not consider in this work is to model the erosion (or change in porosity) of the membrane over time, as discussed recently for a rectangular lattice system \cite{gomes-filho_2022}, where diffusing particles are released from the system only when they reach pores that randomly form over time at lattice sites along the membrane boundary.}

In this work, inspired by our recent work on parameterising, characterising and homogenising continuum and stochastic models of diffusion \cite{carr_2019d,carr_2019a,carr_2019e,carr_2018c}, we propose a moment matching approach to develop several simple one-term models for $\mathcal{P}(t)$. Our approach involves assuming a functional form for $\mathcal{P}(t)$ with parameters identified by matching moments with $\mathcal{P}_{\mathrm{c}}(t)$, the continuum representation of $\mathcal{P}(t)$ obtained from the continuum analogue of the stochastic diffusion model. First, we explore approximating $\mathcal{P}(t)$ by a one-parameter exponential function whose zeroth moment matches with the zeroth moment of $\mathcal{P}_{c}(t)$. Second, we explore approximating $\mathcal{P}(t)$ by a two-parameter Weibull function whose zeroth and first moments match with the zeroth and first moments of $\mathcal{P}_{\mathrm{c}}(t)$. In total, three distinct problems are considered (i) circular and spherical systems with an absorbing boundary (ii) circular and spherical systems with a semi-absorbing boundary and (iii) annular and spherical-shell systems with absorbing, reflecting or semi-absorbing inner and outer boundaries. For each problem, our approach yields exponential and Weibull models for $\mathcal{P}(t)$ that depend explicitly on the dimension, diffusivity, geometry and boundary conditions of the diffusive transport system and agree well with values of $\mathcal{P}(t)$ obtained from stochastic model simulations and continuum model solutions.

The remaining sections of the paper are structured as follows. First, we state the stochastic and continuum models of diffusion considered in this work and describe how $\mathcal{P}(t)$ is calculated in each model (Sections \ref{sec:stochastic_model} and \ref{sec:continuum_model}). Second, we develop exponential and Weibull models for $\mathcal{P}(t)$ for each of the three distinct problems mentioned above (Sections \ref{sec:exponential_model} and \ref{sec:weibull_model}). Third, we perform computational experiments to assess the accuracy of the developed exponential and Weibull models (Section \ref{sec:results}). Finally, we summarise the work and outline avenues for possible future research (Section \ref{sec:conclusions}).

\section{Stochastic model}
\label{sec:stochastic_model}
Consider $N_{p}$ non-interacting particles, initially uniformly-distributed and  undergoing a lattice-free random walk in a $d$-dimensional domain $\Omega_{d} = \{\mathbf{x}\in\mathbb{R}^{d}\,|\,\ell_{0} < \|\mathbf{x}\| < \ell_{1}\}$, where $\|\mathbf{x}\|$ is the Euclidean norm. Let $\mathbf{x}_{j}(t)\in\Omega_{d}$ be the position of the $j$th particle at time $t$ with initial position 
\begin{gather*}
\mathbf{x}_{j}(0) = \begin{cases} r_{j}, & \text{if $d=1$},\\ r_{j}\,[\cos(\theta_{j}),\sin(\theta_{j})], & \text{if $d=2$},\\ r_{j}\,[\sin(\phi_{j})\cos(\theta_{j}),\sin(\phi_{j})\sin(\theta_{j}),\cos(\phi_{j})], & \text{if $d=3$},\end{cases}  
\end{gather*}
where $\theta_{j} \sim \mathcal{U}[0,2\pi]$, $\phi_{j} = \cos^{-1}(1-2u_{j})$ and \revision{$r_{j} = (v_{j}(\ell_{1}^{d}-\ell_{0}^{d}) + \ell_{0}^{d})^{1/d}$} with $u_{j}\sim \mathcal{U}[0,1]$ and $v_{j}\sim \mathcal{U}[0,1]$. Note the formulas for $\phi_{j}$ and $r_{j}$ ensure that the particles are initially uniformly distributed by area ($d=2$) and volume ($d=3$) avoiding the clustering of points that occur when naively taking $\phi_{j}\sim \mathcal{U}[0,\pi]$ and $r_{j}\in \mathcal{U}[\ell_{0},\ell_{1}]$ \cite{carr_2020b}. Each particle undergoes a random walk with constant steps of distance $\delta > 0$ and duration $\tau>0$, such that during the time step from $t$ to $t+\tau$, each particle either moves to a new position:
\begin{gather*}
\mathbf{x}_{j}(t + \tau) = \mathbf{x}_{j}(t) + \begin{cases} \delta\,\text{sign}(u_{j}-0.5), & \text{if $d=1$},\\ \delta\,[\cos(\theta_{j}),\sin(\theta_{j})], & \text{if $d=2$},\\ \delta\,[\sin(\phi_{j})\cos(\theta_{j}),\sin(\phi_{j})\sin(\theta_{j}),\cos(\phi_{j})], & \text{if $d=3$},\end{cases}  
\end{gather*}
with probability $P$ or remains at its current position, $\mathbf{x}_{j}(t+\tau) = \mathbf{x}_{j}(t)$, with probability $1-P$. Here $u_{j}\sim\mathcal{U}[0,1]$, $\theta_{j} \sim \mathcal{U}[0,2\pi]$ and $\phi_{j} = \cos^{-1}(1-2u_{j})$ as above. If during the time step from $t$ to $t+\tau$ movement of the $j$th particle requires it to (i) pass through an absorbing boundary, then the particle is removed from the system, (ii) pass through a reflecting boundary, then the particle remains at its current position, $\mathbf{x}_{j}(t+\tau) = \mathbf{x}_{j}(t)$ (iii) pass through a semi-absorbing boundary at $x = \ell_{0}$, then the particle is removed from the system with probability $P_{0}$ and remains at its current position with probability $1-P_{0}$ (iv) pass through a semi-absorbing boundary at $x = \ell_{1}$, then the particle is removed from the system with probability $P_{1}$ and remains at its current position with probability $1-P_{1}$. 

For the stochastic model, $\mathcal{P}(t)$ is defined as \cite{gomes-filho_2020}
\begin{gather}
\label{eq:Ps}
\mathcal{P}_{\mathrm{s}}(t) = \frac{N(t)}{N_{p}},
\end{gather}
where $N(t)$ is the number of particles remaining in the system at time $t$. Note that $\mathcal{P}_{\mathrm{s}}(0) = 1$ and $\lim_{t\rightarrow\infty}\mathcal{P}_{\mathrm{s}}(t) = 0$ as eventually all particles exit the system when there is at least one absorbing or semi-absorbing boundary (as is evident in Figure \ref{fig:figure1}).

\section{Continuum model}
\label{sec:continuum_model}
The continuum analogue of the stochastic model, outlined in the previous section, is the $d$-dimensional radially-symmetric diffusion equation \cite{codling_2008,okubo_2001,ibe_2013,vaccario_2015} for the dimensionless particle concentration $c(r,t)$:
\begin{gather}
\label{eq:cm_pde}
\frac{\partial c}{\partial t} = \frac{D}{r^{d-1}}\frac{\partial}{\partial r}\left(r^{d-1}\frac{\partial c}{\partial r}\right),
\end{gather}
where $D = P\delta^{2}/(2d\tau)$ is the diffusivity. Here, $c(r,t) = \widetilde{c}(r,t)/\widetilde{c}_{0}$ where $\widetilde{c}(r,t)$ is the particle concentration, which is initially uniform, $\widetilde{c}(r,0) = \widetilde{c}_{0}$. The appropriate initial and boundary conditions are
\begin{gather}
\label{eq:cm_ic}
c(r,0) = 1,\\ 
\label{eq:cm_bc1}
a_{0}c(\ell_{0},t) - b_{0}\frac{\partial c}{\partial r}(\ell_{0},t) = 0,\\
\label{eq:cm_bc2}
a_{1}c(\ell_{1},t) + b_{1}\frac{\partial c}{\partial r}(\ell_{1},t) = 0,
\end{gather}
where the coefficients depend on whether the inner and outer boundaries are designated as absorbing, reflecting or semi-absorbing:
\begin{align}
\label{eq:sm_bc1}
[a_{0},b_{0}] &= \begin{cases} [1,0], & \text{if the inner boundary is absorbing},\\
[0,1], & \text{if the inner boundary is reflecting},\\
[1,\sigma_{0}], & \text{if the inner boundary is semi-absorbing},
\end{cases}\\
\label{eq:sm_bc2}
[a_{1},b_{1}] &= \begin{cases} [1,0], & \text{if the outer boundary is absorbing},\\
[0,1], & \text{if the outer boundary is reflecting},\\
[1,\sigma_{1}], & \text{if the outer boundary is semi-absorbing},
\end{cases}
\end{align}
with $\sigma_{0} = \delta/P_{0}$ and $\sigma_{1} = \delta/P_{1}$. Note that the continuum model is derived in the continuum limit as $\delta\rightarrow 0$ and $\tau\rightarrow 0$ and therefore is valid in the regime of small $\delta$ and $\tau$ \cite{okubo_2001,codling_2008}.

The coefficients for absorbing and reflecting boundaries in (\ref{eq:sm_bc1}) and (\ref{eq:sm_bc2}) are standard while the values for semi-absorbing boundaries can be obtained by considering the probability a particle is located at the inner or outer boundary at a given point in time \cite{erban_2007}. For example, for $d = 1$, if a particle is located at the outer boundary ($x=\ell_{1}$) at time $t + \tau$, then there are three possibilities for its location at time $t$ (i) it was located at $x = \ell_{1}-\delta$ and it moved right (occurs with probability $\frac{P}{2}$) (ii) it was located at $x = \ell_{1}$ and it didn't attempt to move (occurs with probability $1-P$) (iii) it was located at $x = \ell_{1}$, attempted to move right but was reflected (occurs with probability $\frac{P}{2}(1-P_{1})$). Hence, if $p(r,t)$ is the probability a particle is located at position $r$ at time $t$ then $p(\ell_{1},t+\tau) = \frac{P}{2}p(\ell_{1}-\delta,t) + (1-P+\frac{P}{2}(1-P_{1}))p(\ell_{1},t)$. Expanding this equation in a Taylor series yields $p(\ell_{1},t) + \frac{\delta}{P_{1}}\frac{\partial p}{\partial x}(\ell_{1},t) = 0$ when neglecting higher-order terms and ultimately equations (\ref{eq:cm_bc2}) and (\ref{eq:sm_bc2}) when multiplying by $N_{p}$ and using the relationship $c(r,t)=N_{p}p(r,t)$. Similar arguments can be made for $d=2$ and $d=3$. 

For the continuum model, $\mathcal{P}(t)$ is defined as \cite{ignacio_2021}
\begin{gather*}
\mathcal{P}_{\mathrm{c}}(t) = \frac{\int_{\Omega_{d}} c(r,t)\,\text{d}V}{\int_{\Omega_{d}} c(r,0)\,\text{d}V}.
\end{gather*}
Using the initial condition (\ref{eq:cm_ic}), $\mathcal{P}_{\mathrm{c}}(t)$ simplifies to the spatial average of $c(r,t)$ over $\Omega_{d}$
\begin{gather*} 
\mathcal{P}_{\mathrm{c}}(t) = \frac{1}{|\Omega_{d}|}{\int_{\Omega_{d}} c(r,t)\,\text{d}V},
\end{gather*}
where $|\Omega_{d}|$ is the length ($d=1$), area ($d=2$) or volume ($d=3$) of $\Omega_{d}$. Finally, using the definition of the integral in Cartesian, polar and spherical coordinates and the value of $|\Omega_{d}|$ we obtain
\begin{gather} 
\label{eq:Pc}
\mathcal{P}_{\mathrm{c}}(t) = \frac{d}{\ell_{1}^{d}-\ell_{0}^{d}}\int_{\ell_{0}}^{\ell_{1}} r^{d-1}c(r,t)\,\text{d}r,
\end{gather}
due to radial symmetry. Consistent with the stochastic model, $\mathcal{P}_{\mathrm{c}}(0) = 1$ and $\lim_{t\rightarrow\infty}\mathcal{P}_{\mathrm{c}}(t) = 0$ when there is at least one absorbing or semi-absorbing boundary.

Here, we acknowledge that exact expressions for $\mathcal{P}_{\mathrm{c}}(t)$ can be obtained by solving the continuum model (\ref{eq:cm_pde})--(\ref{eq:cm_bc2}) for $c(r,t)$ using separation of variables and eigenfunction expansion and then applying the averaging operator (\ref{eq:Pc}). However, as pointed out in the Introduction, such expressions take the form of an infinite series of exponential functions that make fitting and interpreting experimental release data difficult \cite{ignacio_2017}. As an alternative way forward, one could also think about applying the averaging operator (\ref{eq:Pc}) to the diffusion equation (\ref{eq:cm_pde}) to obtain a differential equation satisfied by $\mathcal{P}_{\mathrm{c}}(t)$, however, as we show in Appendix \ref{sec:averaging} this is only possible for the special case of reflecting boundary conditions at $x = \ell_{0}$ and $x=\ell_{1}$, where trivially $\mathcal{P}_{c}(t) = 1$ for all time as no particles exit the system. To address these issues, in the following sections, we develop several one-term exponential and Weibull models to approximate $\mathcal{P}_{\mathrm{c}}(t)$ and hence $\mathcal{P}(t)$.

\section{Exponential model}
\label{sec:exponential_model} 
We first consider approximating $\mathcal{P}(t)$ using the exponential model:
\begin{gather}
\label{eq:exponential_model}
\mathcal{P}_{\mathrm{e}}(t) = e^{-Dt/\lambda},
\end{gather}
where $\lambda > 0$ is a constant that depends on the dimension ($d$), geometrical lengths ($\ell_{0}$, $\ell_{1}$) and boundary coefficients ($a_{0},a_{1},b_{0},b_{1}$) but not on the diffusivity ($D$). Note that $\mathcal{P}_{\mathrm{e}}(t)$ agrees with $\mathcal{P}_{\mathrm{c}}(t)$ initially at $t=0$ and in the long time limit as $t\rightarrow\infty$. The value of $\lambda$ is chosen to match the zeroth moments of $\mathcal{P}_{\mathrm{e}}(t)$ and $\mathcal{P}_{\mathrm{c}}(t)$:
\begin{gather}
\label{eq:constraint}
\int_{0}^{\infty} \mathcal{P}_{\mathrm{e}}(t)\,\text{d}t = \int_{0}^{\infty} \mathcal{P}_{\mathrm{c}}(t) \,\text{d}t,
\end{gather}
which yields the formula
\begin{gather}
\label{eq:lambda1}
\lambda = D\int_{0}^{\infty} \mathcal{P}_{\mathrm{c}}(t) \,\text{d}t,
\end{gather}
when inserting the form of $\mathcal{P}_{\mathrm{e}}(t)$ (\ref{eq:exponential_model}). This choice for $\lambda$ is attractive as the resulting exponential model $\mathcal{P}_{\mathrm{e}}(t)$ (\ref{eq:exponential_model}) agrees well with $\mathcal{P}_{\mathrm{c}}(t)$ (see Section \ref{sec:results}) and because it allows closed-form analytical expressions for $\lambda$ to be derived without identifying $c(x,t)$ as we now show. Inserting $\mathcal{P}_{\mathrm{c}}(t)$ (\ref{eq:Pc}) into the formula for $\lambda$ (\ref{eq:lambda1}) and interchanging the order of integration gives
\begin{gather}
\label{eq:lambda}
\lambda = D\langle M\rangle,
\end{gather}
where
\begin{gather}
\label{eq:Mavg}
\langle{M}\rangle = \frac{d}{\ell_{1}^{d}-\ell_{0}^{d}}\int_{\ell_{0}}^{\ell_{1}} r^{d-1}M(r)\,\text{d}r,\\
\label{eq:M}
M(r) = \int_{0}^{\infty} c(r,t)\,\text{d}t.
\end{gather}
The attraction here is that $\langle M\rangle$ (and hence $\lambda$) can be calculated explicitly since $M(r)$ satisfies a boundary value problem with a simple closed-form solution \citep{carr_2019e}. The appropriate differential equation is obtained by noting that
\begin{align*}
\frac{D}{r^{d-1}}\frac{\text{d}}{\text{d}r}\left(r^{d-1}\frac{\text{d}M}{\text{d}r}\right) = \int_{0}^{\infty}\frac{D}{r^{d-1}}\frac{\partial}{\partial r}\left(r^{d-1}\frac{\partial c}{\partial r}\right)\,\text{d}t = \int_{0}^{\infty}\frac{\partial c}{\partial t}\,\text{d}t = -1,
\end{align*}
while the appropriate boundary conditions are formulated by combining the integral definition of $M(r)$ with the boundary conditions for $c(r,t)$ (\ref{eq:cm_bc1})--(\ref{eq:cm_bc2}). In summary, $M(r)$ satisfies the following boundary value problem:
\begin{gather}
\label{eq:M_ode}
\frac{D}{r^{d-1}}\frac{\text{d}}{\text{d}r}\left(r^{d-1}\frac{\text{d}M}{\text{d}r}\right) = -1,\\
\label{eq:M_bc1}
a_{0}M(\ell_{0}) - b_{0}\frac{\text{d}M}{\text{d}r}(\ell_{0}) = 0,\\
\label{eq:M_bc2}
a_{1}M(\ell_{1}) + b_{1}\frac{\text{d}M}{\text{d}r}(\ell_{1}) = 0.
\end{gather}
\revision{In the context of first passage processes, $M(r)$ is precisely the mean exit time (mean time required to be released) for a particle starting at a distance $r$ from the origin and (\ref{eq:M_ode})--(\ref{eq:M_bc2}) is the well-known boundary value problem for determining the mean exit time \cite{redner_2001}. It follows then that applying (\ref{eq:constraint}) yields the value of $\lambda$ that produces the same mean exit time averaged across all starting locations in~$\Omega_{d}$.}

In the following sections, we state several exponential models for $\mathcal{P}(t)$ obtained by solving the boundary value problem (\ref{eq:M_ode})--(\ref{eq:M_bc2}) for $M(r)$ and computing $\lambda$ according to equations (\ref{eq:lambda}) and (\ref{eq:Mavg}). Three distinct cases are considered spanning both circular/spherical systems and annular/spherical-shell systems with at least one absorbing or semi-absorbing boundary. As we will see, in each case, our moment matching approach yields an exponential model that depends explicitly on the dimension ($d$), geometry ($\ell_{0}$, $\ell_{1}$), diffusivity ($D$) and boundary conditions ($a_{0},a_{1},b_{0},b_{1}$). 

\subsection{Circular and spherical systems with absorbing boundary}
\label{sec:exponential_circle_absorbing}
For the case of a circular or spherical system ($\ell_{0} = 0$, $\ell_{1} = L$) with radial symmetry at the origin, $[a_{0},b_{0}] = [0,1]$, and an absorbing outer boundary, $[a_{1},b_{1}] = [1,0]$, we obtain 
\begin{gather*}
\lambda = \frac{L^{2}}{d(d+2)},
\end{gather*}
and hence the following exponential model for $\mathcal{P}(t)$:
\begin{gather}
\label{eq:Pe1}
\mathcal{P}_{\mathrm{e}}(t) = \exp\left(\frac{-d(d+2)Dt}{L^{2}}\right).
\end{gather}
We see that $\mathcal{P}(t)$ decreases more rapidly when increasing $d$, increasing $D$ or decreasing $L$, which makes sense as more particles are released from the system during a given time interval when the system undergoes such changes.

\subsection{Circular and spherical systems with semi-absorbing boundary}
\label{sec:exponential_circle_semi}
For the case of a circular or spherical system ($\ell_{0} = 0$, $\ell_{1} = L$) with radial symmetry at the origin, $[a_{0},b_{0}] = [0,1]$, and a semi-absorbing outer boundary, $[a_{1},b_{1}] = [1,\sigma_{1}]$, we obtain 
\begin{gather*}
\lambda = \frac{L^{2}+\sigma_{1}(d+2)L}{d(d+2)},
\end{gather*}
and hence the following exponential model for $\mathcal{P}(t)$:
\begin{gather}
\label{eq:Pe2}
\mathcal{P}_{\mathrm{e}}(t) = \exp\left(\frac{-d(d+2)Dt}{L^{2} + \sigma_{1}(d+2)L}\right).
\end{gather}
Note that (\ref{eq:Pe1}) is recovered when $\sigma_{1} = 0$. Since $\sigma_{1}(d+2)L > 0$, we also see that $\mathcal{P}(t)$ decreases slower when the outer boundary is semi-absorbing compared to when it is absorbing (\ref{eq:Pe1}), which makes sense since fewer particles are released from the system during a given time interval.
 
\subsection{Annular and spherical-shell systems}
\label{sec:exponential_annulus}
For the case of an annulus or spherical-shell system ($\ell_{0} > 0$), we obtain the following exponential model for $\mathcal{P}(t)$:
\begin{align*}
&\mathcal{P}_{\mathrm{e}}(t) = e^{-Dt/\lambda},\\
\intertext{where $\lambda$ is defined explicitly as follows}
&I_{1} = \int_{\ell_{0}}^{\ell_{1}}r^{1-d}\,\text{d}r,\quad I_{2} = \int_{\ell_{0}}^{\ell_{1}}r^{d-1}\!\int_{\ell_{0}}^{r}s^{1-d}\,\text{d}s\,\text{d}r,\\
&\beta_{1} = \frac{(a_{1}I_{1} + b_{1}\ell_{1}^{1-d})(a_{0}\ell_{0}^{2} - 2b_{0}\ell_{0}) + b_{0}\ell_{0}^{1-d}(a_{1}\ell_{1}^{2}+2b_{1}\ell_{1})}{a_{0}(a_{1}I_{1} + b_{1}\ell_{1}^{1-d}) + a_{1}b_{0}\ell_{0}^{1-d}},\\
&\beta_{2} = \frac{a_{0}a_{1}(\ell_{1}^{2}-\ell_{0}^{2}) + 2(a_{0}b_{1}\ell_{1} + a_{1}b_{0}\ell_{0})}{a_{0}(a_{1}I_{1} + b_{1}\ell_{1}^{1-d}) + a_{1}b_{0}\ell_{0}^{1-d}},\\
&\lambda = \frac{\beta_{1}(d+2)(\ell_{1}^{d}-\ell_{0}^{d})+\beta_{2}d(d+2)I_{2}-d(\ell_{1}^{d+2}-\ell_{0}^{d+2})}{2d(d+2)(\ell_{1}^{d}-\ell_{0}^{d})},
\end{align*}
and $a_{0}$, $b_{0}$, $a_{1}$, $b_{1}$ specify whether the inner and outer boundaries are absorbing, reflecting or semi-absorbing as per the definitions (\ref{eq:sm_bc1})--(\ref{eq:sm_bc2}). Note that $\lambda$ depends on two geometrical constants, $I_{1}$ and $I_{2}$, which are expressed in terms of definite integrals for succinctness. Closed-form expressions for these integrals are given in Appendix \ref{sec:integrals} for $d = 1,2,3$.

\section{Weibull model}
\label{sec:weibull_model} 
We now consider approximating $\mathcal{P}(t)$ using the Weibull function
\begin{gather}
\label{eq:weibull_model}
\mathcal{P}_{\mathrm{w}}(t) = e^{-(Dt/\mu)^\alpha},
\end{gather}
where $\mu > 0$ and $\alpha >0 $ (the latter typically in the range $[0.5,1]$ \cite{ignacio_2021}) are constants that depend on the dimension ($d$), geometrical lengths ($\ell_{0}$, $\ell_{1}$) and boundary coefficients ($a_{0},a_{1},b_{0},b_{1}$) but not on the diffusivity ($D$). In a similar manner to the exponential model, $\mathcal{P}_{\mathrm{w}}(t)$ agrees with $\mathcal{P}_{\mathrm{c}}(t)$ initially at $t = 0$ and in the long time limit $t\rightarrow\infty$. The values of $\mu$ and $\alpha$ are chosen to match the zeroth and first moments of $\mathcal{P}_{\mathrm{w}}(t)$ and $\mathcal{P}_{\mathrm{c}}(t)$:
\begin{gather}
\label{eq:constraint1}
\int_{0}^{\infty} \mathcal{P}_{\mathrm{w}}(t)\,\text{d}t = \int_{0}^{\infty} \mathcal{P}_{\mathrm{c}}(t) \,\text{d}t,\\
\label{eq:constraint2}
\int_{0}^{\infty} t\mathcal{P}_{\mathrm{w}}(t)\,\text{d}t = \int_{0}^{\infty} t\mathcal{P}_{\mathrm{c}}(t) \,\text{d}t,
\end{gather}
which yields the following constraints on $\mu$ and $\alpha$:
\begin{align}
\label{eq:int_reduced_order1}
\frac{\mu\Gamma(\frac{1}{\alpha})}{\alpha} &= \int_{0}^{\infty} \mathcal{P}_{\mathrm{c}}(t)\,\text{d}t,\\
\label{eq:int_reduced_order2}
\frac{\mu^2\Gamma(\frac{2}{\alpha})}{\alpha} &= \int_{0}^{\infty} t\mathcal{P}_{\mathrm{c}}(t)\,\text{d}t,
\end{align}
when inserting the form of $\mathcal{P}_{\mathrm{w}}(t)$ (\ref{eq:exponential_model}). Inserting $\mathcal{P}_{\mathrm{c}}(t)$ (\ref{eq:Pc}) into these equations and interchanging the order of integration yields a pair of coupled nonlinear equations:
\begin{align}
\label{eq:weibull_n1}
\frac{\mu\Gamma(\frac{1}{\alpha})}{\alpha} &= D\langle{M}_{1}\rangle,\\
\label{eq:weibull_n2}
\frac{\mu^2\Gamma(\frac{2}{\alpha})}{\alpha} &= D^{2}\langle{M}_{2}\rangle,
\end{align}
where 
\begin{gather}
\label{eq:M12avg}
\langle{M}_{1}\rangle = \frac{d}{\ell_{1}^{d}-\ell_{0}^{d}}\int_{\ell_{0}}^{\ell_{1}} r^{d-1}M_{1}(r)\,\text{d}r,\quad
\langle{M}_{2}\rangle = \frac{d}{\ell_{1}^{d}-\ell_{0}^{d}}\int_{\ell_{0}}^{\ell_{1}} r^{d-1}M_{2}(r)\,\text{d}r,\\
\label{eq:M12}
M_{1}(r) = \int_{0}^{\infty} c(r,t)\,\text{d}t,\quad
M_{2}(r) = \int_{0}^{\infty} tc(r,t)\,\text{d}t.
\end{gather}
As for the exponential model, the attraction here is that both $\langle M_{1}\rangle$ and $\langle M_{2}\rangle$ can be calculated explicitly since $M_{1}(r)$ and $M_{2}(r)$ satisfy boundary value problems with closed-form solutions. The function $M_{1}(r)$ is equivalent to $M(r)$ from the exponential model (\ref{eq:M}) so it satisfies the boundary value problem (\ref{eq:M_ode})--(\ref{eq:M_bc2}). On the other hand, the appropriate differential equation for $M_{2}(r)$ is obtained by noting that:
\begin{align*}
\frac{D}{r^{d-1}}\frac{\text{d}}{\text{d}r}\left(r^{d-1}\frac{\text{d}M_{2}}{\text{d}r}\right) = \int_{0}^{\infty} t \frac{D}{r^{d-1}}\frac{\partial}{\partial r}\left(r^{d-1}\frac{\partial c}{\partial r}\right)\,\text{d}t = \int_{0}^{\infty}t\frac{\partial c}{\partial t}\,\text{d}t = -\int_{0}^{\infty} c(r,t)\,\text{d}t = -M_{1}(r),
\end{align*}
with boundary conditions formulated by combining the integral definition of $M_{2}(r)$ (\ref{eq:M12}) with the boundary conditions for $c(r,t)$ (\ref{eq:cm_bc1})--(\ref{eq:cm_bc2}). In summary, $M_{1}(r)$ satisfies the boundary value problem:
\begin{gather}
\label{eq:M1_ode}
\frac{D}{r^{d-1}}\frac{\text{d}}{\text{d}r}\left(r^{d-1}\frac{\text{d}M_{1}}{\text{d}r}\right) = -1,\\
\label{eq:M1_bc1}
a_{0}M_{1}(\ell_{0}) - b_{0}\frac{\text{d}M_{1}}{\text{d}r}(\ell_{0}) = 0,\\
\label{eq:M1_bc2}
a_{1}M_{1}(\ell_{1}) + b_{1}\frac{\text{d}M_{1}}{\text{d}r}(\ell_{1}) = 0,
\end{gather}
and $M_{2}(r)$ satisfies the boundary value problem:
\begin{gather}
\label{eq:M2_ode}
\frac{D}{r^{d-1}}\frac{\text{d}}{\text{d}r}\left(r^{d-1}\frac{\text{d}M_{2}}{\text{d}r}\right) = -M_{1},\\
\label{eq:M2_bc1}
a_{0}M_{2}(\ell_{0}) - b_{0}\frac{\text{d}M_{2}}{\text{d}r}(\ell_{0}) = 0,\\
\label{eq:M2_bc2}
a_{1}M_{2}(\ell_{1}) + b_{1}\frac{\text{d}M_{2}}{\text{d}r}(\ell_{1}) = 0.
\end{gather}
\revision{In the context of first passage processes, $M_{1}(r)$ and $M_{2}(r)$ are precisely the first (mean) and second moments of exit time for a particle starting at a distance $r$ from the origin while (\ref{eq:M1_ode})--(\ref{eq:M1_bc2}) and (\ref{eq:M2_ode})--(\ref{eq:M2_bc2}) are the well-known boundary value problems for determining these moments \cite{redner_2001}. It follows then that applying (\ref{eq:constraint1}) and (\ref{eq:constraint2}) yields the values of $\mu$ and $\alpha$ that produce the same first and second moments of exit time averaged across all starting locations in $\Omega_{d}$.}

\begin{figure}
\centering
\includegraphics[height=0.4\textwidth]{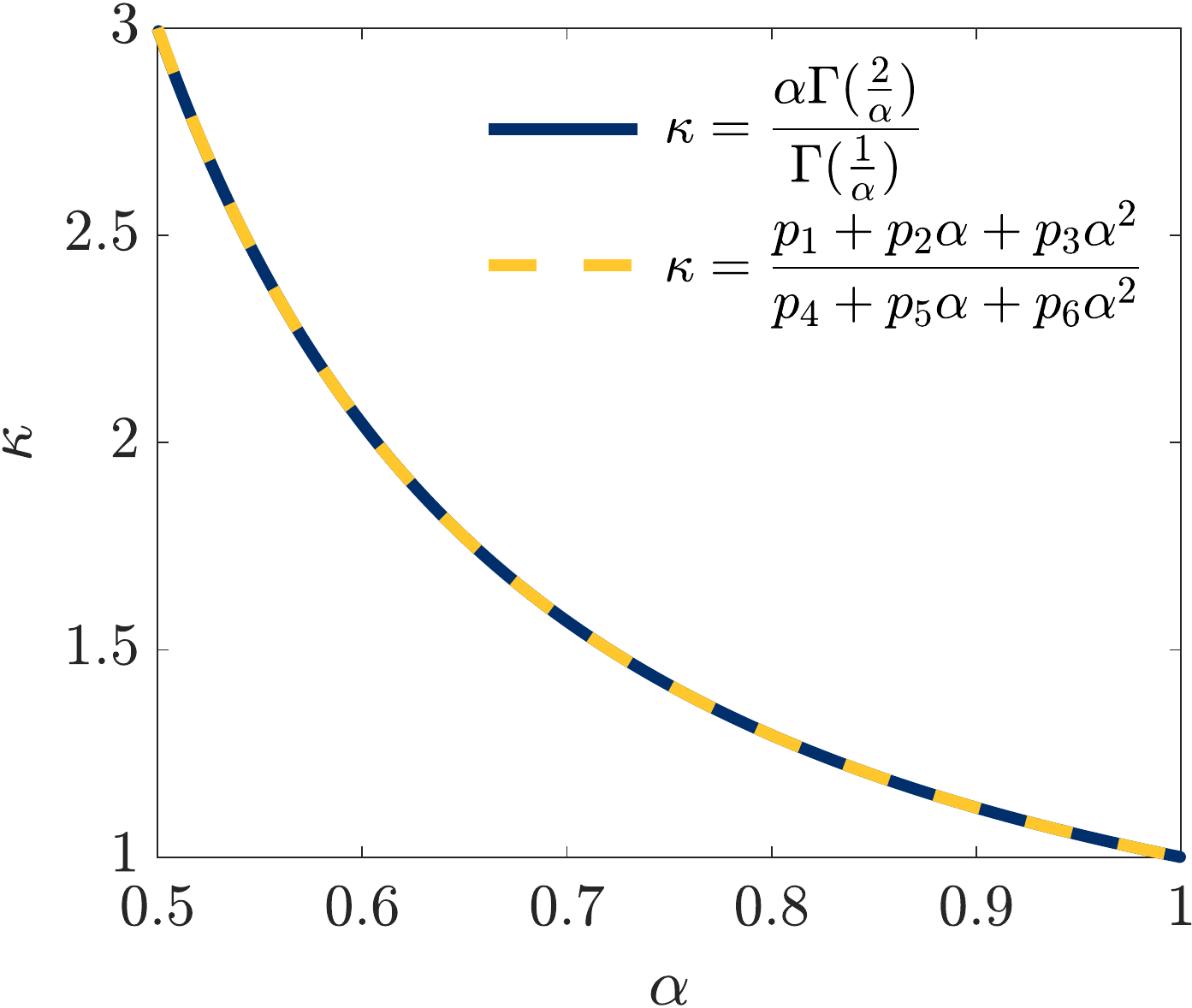}
\caption{The function of $\alpha$ featuring in the nonlinear equation (\ref{eq:alpha_nonlinear}) is closely approximated by the (2,2) Pad\'e approximation featuring in the nonlinear equation (\ref{eq:alpha_pade}). The maximum absolute difference between these two functions on $[0.5,1]$ is approximately $1.3\times 10^{-3}$.}
\label{fig:pade}
\end{figure}

Given closed-form expressions for $\langle M_{1}\rangle$ and $\langle M_{2}\rangle$, the question still remains of how to obtain explicit expressions for $\mu$ and $\alpha$ satisfying the coupled nonlinear equations (\ref{eq:weibull_n1})--(\ref{eq:weibull_n2}). Dividing (\ref{eq:weibull_n2}) by the square of (\ref{eq:weibull_n1}) we see that $\alpha$ satisfies a nonlinear equation that is independent of $\mu$:
\begin{align}
\label{eq:alpha_nonlinear}
\frac{\alpha\Gamma(\frac{2}{\alpha})}{\Gamma(\frac{1}{\alpha})^2} = \kappa,
\end{align}
where $\kappa := \langle M_{2}\rangle/\langle M_{1}\rangle^{2}$. This nonlinear equation maps each value of $\kappa\in[1,3]$ to a unique value of $\alpha\in[0.5,1]$ implicitly (as evident in Figure \ref{fig:pade}). To avoid this implicit relationship, we determine an approximate explicit expression for $\alpha$ by approximating the left-hand side of (\ref{eq:alpha_nonlinear}) by its $(2,2)$ Pad\'e approximation \cite{baker_2014} centered at the midpoint of $[0.5,1]$. This approximation is very accurate for $\alpha\in[0.5,1]$ (as evident in Figure \ref{fig:pade}) and yields the following approximate nonlinear equation:
\begin{align}
\label{eq:alpha_pade}
\frac{p_{1} + p_{2}\alpha + p_{3}\alpha^{2}}{p_{4} + p_{5}\alpha + p_{6}\alpha^{2}} = \kappa,
\end{align}
where $p_{1}=0.45810$, $p_{2}=0.15757$, $p_{3} = 1.49126$, $p_{4} = 0.13963$, $p_{5} = -1.31348$ and $p_{6} = 3.28085$ are rounded to five decimal places and computed using Maple's \verb"pade" function \cite{maple_pade}. The attraction of the nonlinear equation (\ref{eq:alpha_pade}) over the nonlinear equation (\ref{eq:alpha_nonlinear}) is that it can be solved exactly as it reduces to the following quadratic equation on $[0.5,1]$:
\begin{align*}
p_{1} - p_{4}\kappa + \left(p_{2}-p_{5}\kappa\right)\alpha + (p_{3}-p_{6}\kappa)\alpha^{2} = 0.
\end{align*}
Applying the quadratic formula and taking the solution in $[0.5,1]$ yields
\begin{align}
\label{eq:alpha}
\alpha = \frac{p_{5}\kappa - p_{2} - \sqrt{(p_{5}\kappa - p_{2})^2 - 4(p_{3}-p_{6}\kappa)(p_{1}-p_{4}\kappa)}}{2(p_{3}-p_{6}\kappa)},
\end{align}
which provides an approximate explicit formula for $\alpha$ in terms of the system parameters since $\kappa$ depends explicitly on these parameters. In fact, using the value of $\alpha$ calculated from (\ref{eq:alpha}) instead of the exact value of $\alpha$ calculated by numerically solving the nonlinear equation (\ref{eq:alpha_nonlinear}) has almost no impact on the performance of the Weibull model, with the two values of $\alpha$ differing by a most $9.2\times 10^{-5}$ (approximately) when $\kappa\in[1,3]$. In conclusion, with an explicit formula for $\alpha$, we complete the parameterisation of the Weibull model by rearranging (\ref{eq:weibull_n1}) to give
\begin{align}
\label{eq:mu}
\mu = \frac{\alpha D\langle M_{1}\rangle}{\Gamma(\frac{1}{\alpha})}.
\end{align} 
In the following sections, we state several Weibull models for $\mathcal{P}(t)$ obtained by solving the boundary value problems (\ref{eq:M1_ode})--(\ref{eq:M1_bc2}) and (\ref{eq:M2_ode})--(\ref{eq:M2_bc2}) for $M_{1}(r)$ and $M_{2}(r)$ and computing $\alpha$ and $\mu$ according to equations (\ref{eq:alpha}) and (\ref{eq:mu}). As for the exponential model, three distinct cases are considered spanning both circular/spherical systems and annular/spherical-shell systems with at least one absorbing or semi-absorbing boundary. As we will see, in each case, our moment matching approach yields a Weibull model for $\mathcal{P}(t)$ that depends explicitly on the dimension ($d$), geometry ($\ell_{0}$, $\ell_{1}$), diffusivity ($D$) and boundary conditions ($a_{0},a_{1},b_{0},b_{1}$).

\subsection{Circular and spherical systems with absorbing boundary}
\label{sec:weibull_circle_absorbing}
For the case of a circular or spherical system ($\ell_{0} = 0$, $\ell_{1} = L$) with radial symmetry at the origin, $[a_{0},b_{0}] = [0,1]$, and an absorbing outer boundary, $[a_{1},b_{1}] = [1,0]$, we obtain the following Weibull model for $\mathcal{P}(t)$: 
\begin{align*}
&\mathcal{P}_{\mathrm{w}}(t) = e^{-(Dt/\mu)^\alpha},
\intertext{where $\alpha$ and $\mu$ are defined explicitly as follows}
&\alpha = \begin{cases} 0.84883, & \text{if $d = 1,$}\\ 0.78258, & \text{if $d = 2,$}\\ 0.74510, & \text{if $d = 3,$} \end{cases}\\
&\mu = \frac{\alpha L^{2}}{d(d+2)\Gamma(1/\alpha)}.
\end{align*}
For this special case, $\kappa = 2(d+2)/(d+4)$ in the nonlinear equation (\ref{eq:alpha_nonlinear}) and hence $\alpha$ depends on the dimension ($d$) only. Directly solving this resulting nonlinear equation numerically and rounding the results to five decimal places yields the specified values of $\alpha$ above. 

\subsection{Circular and spherical systems with semi-absorbing boundary}
\label{sec:weibull_circle_semi}
For the case of a circular or spherical system  ($\ell_{0} = 0$, $\ell_{1} = L$) with radial symmetry at the origin, $[a_{0},b_{0}] = [0,1]$, and a semi-absorbing outer boundary, $[a_{1},b_{1}] = [1,\sigma_{1}]$, we obtain the following Weibull model for $\mathcal{P}(t)$:
\begin{align*}
&\mathcal{P}_{\mathrm{w}}(t) = e^{-(Dt/\mu)^\alpha},
\intertext{where $\alpha$ and $\mu$ are defined explicitly as follows}
&\kappa = \frac{(d+2)[2L^{4} +\sigma_{1}(d+4)(2L^3+\sigma_{1}(d+2)L^{2})]}{(d+4)(L^{2}+\sigma_{1}(d+2)L)^{2}},\\
&\alpha = \frac{p_{5}\kappa - p_{2} - \sqrt{(p_{5}\kappa - p_{2})^2 - 4(p_{3}-p_{6}\kappa)(p_{1}-p_{4}\kappa)}}{2(p_{3}-p_{6}\kappa)},\\
&\mu = \frac{\alpha[L^{2}+\sigma_{1}(d+2)L]}{d(d+2)\Gamma(1/\alpha)},
\end{align*}
and $p_{1}$, $p_{2}$, $p_{3}$, $p_{4}$, $p_{5}$ and $p_{6}$ are the coefficients of the Pad\'e approximation used in equation (\ref{eq:alpha_pade}). Note that if $\alpha = 1$ the corresponding exponential model in Section \ref{sec:exponential_circle_semi} is recovered.

\subsection{Annular and spherical-shell systems}
\label{sec:weibull_annulus}
For the case of an annular or spherical-shell system ($\ell_{0} > 0$), we obtain the following Weibull model for $\mathcal{P}(t)$:
\begin{align*}
&\mathcal{P}_{\mathrm{w}}(t) = e^{-(Dt/\mu)^\alpha},\\
\intertext{where $\mu$ and $\alpha$ are defined explicitly as follows} 
&I_{1} = \int_{\ell_{0}}^{\ell_{1}}r^{1-d}\,\text{d}r,\quad 
I_{2} = \int_{\ell_{0}}^{\ell_{1}}r^{d-1}\!\int_{\ell_{0}}^{r}s^{1-d}\,\text{d}s\,\text{d}r,\\
&I_{3} = \int_{\ell_{0}}^{\ell_{1}}r^{1-d}\!\int_{\ell_{0}}^{r}u^{d-1}\!\int_{\ell_{0}}^{u}s^{1-d}\,\text{d}s\,\text{d}u\,\text{d}r,\quad
I_{4} = \int_{\ell_{0}}^{\ell_{1}}r^{d-1}\!\int_{\ell_{0}}^{r}w^{1-d}\!\int_{\ell_{0}}^{w}u^{d-1}\!\int_{\ell_{0}}^{u}s^{1-d}\,\text{d}s\,\text{d}u\,\text{d}w\,\text{d}r,\\
&\beta_{1} = \frac{(a_{1}I_{1} + b_{1}\ell_{1}^{1-d})(a_{0}\ell_{0}^{2} - 2b_{0}\ell_{0}) + b_{0}\ell_{0}^{1-d}(a_{1}\ell_{1}^{2}+2b_{1}\ell_{1})}{a_{0}(a_{1}I_{1} + b_{1}\ell_{1}^{1-d}) + a_{1}b_{0}\ell_{0}^{1-d}},\\
&\beta_{2} = \frac{a_{0}a_{1}(\ell_{1}^{2}-\ell_{0}^{2}) + 2(a_{0}b_{1}\ell_{1} + a_{1}b_{0}\ell_{0})}{a_{0}(a_{1}I_{1} + b_{1}\ell_{1}^{1-d}) + a_{1}b_{0}\ell_{0}^{1-d}},\\
&\gamma_{1}^{\ast} = -a_{0}\left[\frac{\ell_{0}^{4}}{4(d+2)} - \beta_{1}\frac{\ell_{0}^2}{2d}\right] + b_{0}\left[\frac{\ell_{0}^{3}}{(d+2)} - \beta_{1}\frac{\ell_{0}}{d}\right],\\
&\gamma_{2}^{\ast} = -a_{1}\left[\frac{\ell_{1}^{4}}{4(d+2)} - \beta_{1}\frac{\ell_{1}^2}{2d} - \beta_{2}I_{3}\right]- b_{1}\left[\frac{\ell_{1}^{3}}{(d+2)} - \beta_{1}\frac{\ell_{1}}{d} - \beta_{2}\ell_{1}^{1-d}I_{2}\right],\\
&\gamma_{1} = \frac{(a_{1}I_{1} + b_{1}\ell_{1}^{1-d})\gamma_{1}^{\ast} + b_{0}\ell_{0}^{1-d}\gamma_{2}^{\ast}}{a_{0}(a_{1}I_{1} + b_{1}\ell_{1}^{1-d}) + a_{1}b_{0}\ell_{0}^{1-d}},\quad
\gamma_{2} = \frac{a_{0}\gamma_{2}^{\ast}-a_{1}\gamma_{1}^{\ast}}{a_{0}(a_{1}I_{1} + b_{1}\ell_{1}^{1-d}) + a_{1}b_{0}\ell_{0}^{1-d}},\\
&\kappa = \frac{ 2(\ell_{1}^{d}-\ell_{0}^{d})\Bigl[\frac{\ell_{1}^{d+4}-\ell_{0}^{d+4}}{4(d+2)(d+4)} - \frac{\beta_{1}(\ell_{1}^{d+2}-\ell_{0}^{d+2})}{2d(d+2)} + \frac{\gamma_{1}(\ell_{1}^{d}-\ell_{0}^{d})}{d} - \beta_{2}I_{4} + \gamma_{2}I_{2}\Bigr]}{\Bigl[\frac{\beta_{1}(\ell_{1}^{d}-\ell_{0}^{d})}{d} - \frac{(\ell_{1}^{d+2}-\ell_{0}^{d+2})}{d+2} + \beta_{2}I_{2}\Bigr]^{2}},\\
&\alpha = \frac{p_{5}\kappa - p_{2} - \sqrt{(p_{5}\kappa - p_{2})^2 - 4(p_{3}-p_{6}\kappa)(p_{1}-p_{4}\kappa)}}{2(p_{3}-p_{6}\kappa)},\\
&\mu = \frac{\alpha[\beta_{1}(d+2)(\ell_{1}^{d}-\ell_{0}^{d})+\beta_{2}d(d+2)I_{2}-d(\ell_{1}^{d+2}-\ell_{0}^{d+2})]}{2d(d+2)(\ell_{1}^{d}-\ell_{0}^{d})\Gamma(1/\alpha)}.
\end{align*}
Recall that in the above expressions $a_{0}$, $b_{0}$, $a_{1}$, $b_{1}$ specify whether the inner and outer boundaries are absorbing, reflecting or semi-absorbing (\ref{eq:sm_bc1})--(\ref{eq:sm_bc2}) and $p_{1}$, $p_{2}$, $p_{3}$, $p_{4}$, $p_{5}$ and $p_{6}$ are the coefficients of the Pad\'e approximation used in equation (\ref{eq:alpha_pade}). Note that $\mu$ and $\alpha$ depend on four geometrical constants, $I_{1}$, $I_{2}$, $I_{3}$ and $I_{4}$, which are expressed in terms of definite integrals for succinctness. Closed-form expressions for these integrals are given in Appendix \ref{sec:integrals} for $d = 1,2,3$.

\section{Computational experiments}
\label{sec:results}
We now assess the accuracy of our exponential and Weibull models for $\mathcal{P}(t)$
against benchmark results obtained from the stochastic (\ref{eq:Ps}) and continuum (\ref{eq:Pc}) models. In our computational experiments, we consider the six test cases in Table \ref{tab:test_cases} involving both circular/spherical (Cases A--B) and annular/spherical-shell systems (Cases C--F) subject to various combinations of absorbing, reflecting and semi-absorbing boundary conditions. The appropriate exponential and Weibull models are given in sections \ref{sec:exponential_circle_absorbing} and \ref{sec:weibull_circle_absorbing} for Case A, sections \ref{sec:exponential_circle_semi} and \ref{sec:weibull_circle_semi} for Case B and sections \ref{sec:exponential_annulus} and \ref{sec:weibull_annulus} for Cases C--F. All comparisons are performed over a specified finite time interval $0 < t < T$. To capture the main region of decrease of $\mathcal{P}(t)$ across the different test cases, we take $T = \mu[k\log_{e}(10)]^{1/\alpha}/D$ with $k=2$, which is the value of time satisfying $\mathcal{P}_{\mathrm{w}}(t) = 10^{-k}$ for the appropriate Weibull model. To compute $\mathcal{P}(t)$ for the stochastic model we consider both $N_{p} = 50$ and $N_{p}=500$ particles and perform $N_{s} = 100$ stochastic simulations with $P = \delta = \tau = 1$. To compute $\mathcal{P}(t)$ for the continuum model, we first solve the continuum diffusion model (\ref{eq:cm_pde})--(\ref{eq:cm_bc2}) numerically with $D = P\delta^{2}/(2d\tau) = 1/(2d)$, discretising in space using a finite volume method with $N_{r} = 501$ uniformly spaced nodes and discretising in time using the Crank-Nicolson method with $N_{t} = 10^{4}$ fixed time steps. This process yields numerical approximations to $c(r_{k},t_{i})$ for $k = 1,\hdots,N_{r}$ and $i = 1,\hdots,N_{t}$, where $r_{k} = \ell_{0} + (k-1)(\ell_{1}-\ell_{0})/(N_{r}-1)$ and $t_{i} = iT/N_{t}$. To compute $\mathcal{P}(t)$ we then combine the numerical approximations to $c(r_{k},t_{i})$ for $k = 1,\hdots,N_{r}$ with a Simpson's rule approximation to the spatial average (\ref{eq:Pc}). Given these benchmark results for $\mathcal{P}(t)$ obtained from the continuum model, accuracy of the exponential and Weibull models are quantified using the errors:
\begin{gather*}
\varepsilon_{\mathrm{e}} = \frac{1}{N_{t}}\sum_{i=1}^{N_{t}}|\mathcal{P}_{\mathrm{e}}(t_{i}) - \mathcal{P}_{\mathrm{c}} (t_{i})|,\quad \varepsilon_{\mathrm{w}} = \frac{1}{N_{t}}\sum_{i=1}^{N_{t}}|\mathcal{P}_{\mathrm{w}}(t_{i}) - \mathcal{P}_{\mathrm{c}} (t_{i})|,
\end{gather*}
where subscripts e, w and c are used to distinguish between the exponential, Weibull and continuum models, respectively, as in Sections \ref{sec:continuum_model}--\ref{sec:weibull_model}. Complete implementation details can be found in our supporting MATLAB code available on GitHub: \href{https://github.com/elliotcarr/Carr2022b}{github.com/elliotcarr/Carr2022b}.

\begin{table*}[b]
\small
\def\arraystretch{1.2}
\begin{tabular*}{\textwidth}{@{\extracolsep{\fill}}lccllllll}
\hline
Case & $\ell_{0}$ & $\ell_{1}$ & Inner boundary & Outer boundary & $a_{0}$ & $b_{0}$ & $a_{1}$ & $b_{1}$\\
\hline
A & $0$ & 100 & n/a & absorbing & 0 & 1 & 1 & 0\\
B & $0$ & 100 & n/a & semi-absorbing ($P_{1}=0.2$) & 0 & 1 & 1 & 5\\
C & $50$ & 100 & reflecting & absorbing & 0 & 1 & 1 & 0\\
D & $50$ & 100 & reflecting & semi-absorbing ($P_{1}=0.2$) & 0 & 1 & 1 & 5\\
E & $50$ & 100 & absorbing & absorbing & 1 & 0 & 1 & 0\\
F & $50$ & 100 & semi-absorbing ($P_{0}=0.5$) & semi-absorbing ($P_{1}=0.2$) & 1 & 2 & 1 & 5\\
\hline
\end{tabular*}
\caption{Geometry and boundary parameters for test cases A--F.}
\label{tab:test_cases}
\end{table*}

\begin{figure*}[p]
\def\figw{0.38\textwidth}
\centering
\includegraphics[width=\textwidth]{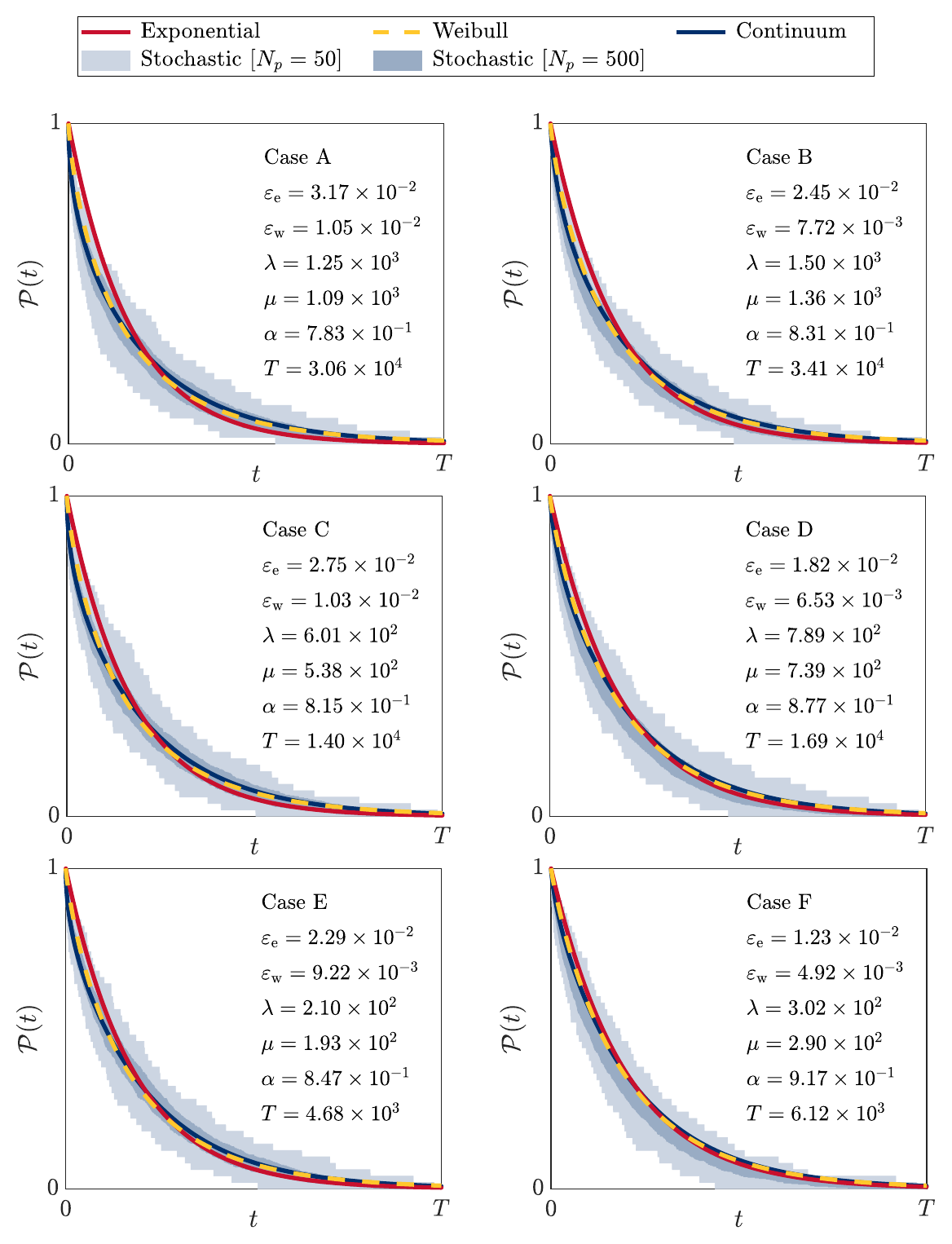}\\
\caption{\textbf{Results for circular and annular systems ($\boldsymbol{d=2}$).} Performance of the Exponential and Weibull models for the proportion of particles remaining over time, $\mathcal{P}(t)$, benchmarked against results for $\mathcal{P}(t)$ obtained from the stochastic (\ref{eq:Ps}) and continuum (\ref{eq:Pc}) models. For the stochastic model, shaded regions represent 95\% confidence intervals (between 2.5\% and 97.5\% quantiles) across the $N_{s}$ stochastic simulations. The exponential model parameter $\lambda$, Weibull model parameters $\mu$ and $\alpha$, errors $\varepsilon_{\mathrm{e}}$ and $\varepsilon_{\mathrm{w}}$ and final time $T$ are all rounded to three significant digits.}
\label{fig:results_2d}
\end{figure*}

\begin{figure*}[p]
\def\figw{0.38\textwidth}
\centering
\includegraphics[width=\textwidth]{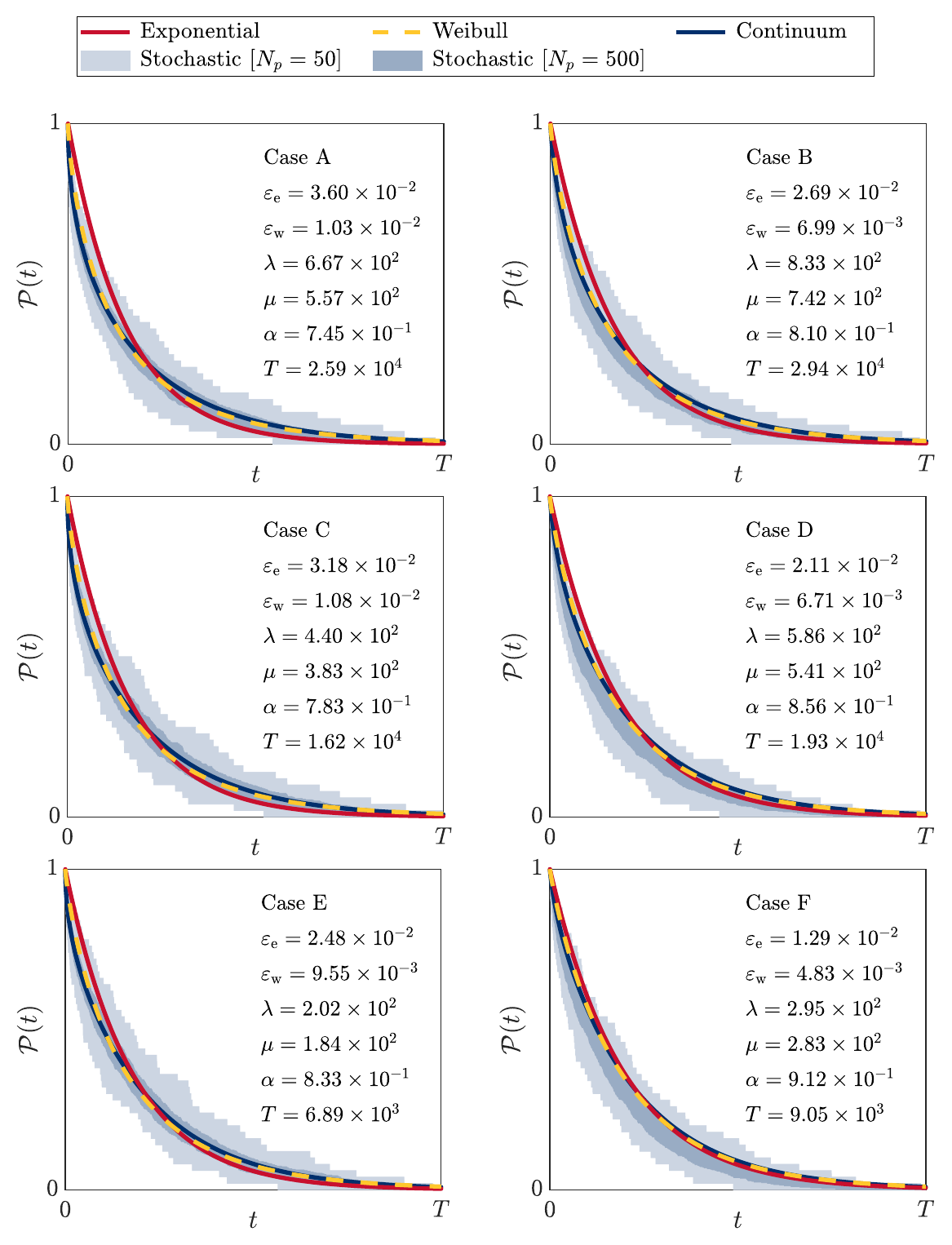}\\
\caption{\textbf{Results for spherical and spherical-shell systems ($\boldsymbol{d=3}$).} Performance of the Exponential and Weibull models for the proportion of particles remaining over time, $\mathcal{P}(t)$, benchmarked against results for $\mathcal{P}(t)$ obtained from the stochastic (\ref{eq:Ps}) and continuum (\ref{eq:Pc}) models. For the stochastic model, shaded regions represent 95\% confidence intervals (between 2.5\% and 97.5\% quantiles) across the $N_{s}$ stochastic simulations. The exponential model parameter $\lambda$, Weibull model parameters $\mu$ and $\alpha$, errors $\varepsilon_{\mathrm{e}}$ and $\varepsilon_{\mathrm{w}}$ and final time $T$ are all rounded to three significant digits.}
\label{fig:results_3d}
\end{figure*}

Results in Figures \ref{fig:results_2d} and \ref{fig:results_3d} plot $\mathcal{P}(t)$ for the exponential, Weibull, continuum and stochastic models for each test case and for dimensions $d = 2,3$. Similar results are obtained for $d=1$ (not shown). Also featured in each plot are the computed values of the exponential model parameter, $\lambda$, the Weibull model parameters, $\mu$ and $\alpha$, the errors, $\varepsilon_{\mathrm{e}}$ and $\varepsilon_{\mathrm{w}}$, and the final time $T$. Note that each plot is on a different time scale as $T$ changes substantially across the test cases. Several observations are evident from these results:
\begin{itemize}
\item Both the exponential and Weibull models reliably capture the release profile and release time-scale across all six test cases.
\item The exponential model provides a lower accuracy (but potentially sufficient) approximation to $\mathcal{P}(t)$ across all six test cases. 
\item The Weibull model provides a higher accuracy approximation to $\mathcal{P}(t)$, more accurately capturing the fast early decay and slow later decay of $\mathcal{P}(t)$, as is especially evident in the test cases with absorbing boundaries (Cases A, C, E).
\item Both the exponential and Weibull models are more accurate for test cases involving semi-absorbing boundary conditions than for test cases involving absorbing boundary conditions (i.e. more accurate for Case B than Case A, Case D than Case C and Case F than Case E). 
\item The value of $T = \mu[k\log_{e}(10)]^{1/\alpha}/D$, involving $\mu$ and $\alpha$ from the Weibull model and satisfying $\mathcal{P}_{\mathrm{w}}(T) = 10^{-k}$, provides a reliable rule-of-thumb approximation of the time when all the particles have been removed from the system (i.e. $\mathcal{P}(t) \approx 0$).
\item The value of $\alpha$ decreases with dimension ($d$), consistent with that reported elsewhere \cite{ignacio_2021}.
\end{itemize}
In summary, both the exponential and Weibull models provide useful ways to characterise $\mathcal{P}(t)$ with the improved accuracy offered by the Weibull model coming at the expense of less elegant formulas.% for $\mathcal{P}(t)$. 

\section{Conclusion}
\label{sec:conclusions}
\noindent We have considered the classical problem of particle diffusion for $d$-dimensional radially-symmetric systems with absorbing and semi-absorbing boundaries, proposing new single-term one and two parameter models for approximating $\mathcal{P}(t)$, the proportion of particles remaining in the system over time. Our approach involves matching moments with the continuum analogue of $\mathcal{P}(t)$ obtained from the continuum limit of the stochastic diffusion model. Exponential and Weibull models for $\mathcal{P}(t)$ were developed for (i) circular and spherical systems with an absorbing boundary (ii) circular and spherical systems with a semi-absorbing boundary and (iii) annular and spherical-shell systems with absorbing, reflecting or semi-absorbing inner and outer boundaries. Results demonstrate that both models reliably capture the profile and time-scale of release, providing easy-to-evaluate approximations of $\mathcal{P}(t)$ that depend explicitly on the dimension, diffusivity, geometry and boundary conditions of the diffusive transport system. Although the Weibull model is more accurate than the exponential model in all reported cases, this comes at the cost of a more complicated formula for $\mathcal{P}(t)$. 

\revision{While the exponential and Weibull models developed in this work approximate $\mathcal{P}(t)$ well at short, long and intermediate times, it is important to note that they do not capture the actual asymptotic form of the short or long-time  behaviour of $\mathcal{P}_{\mathrm{c}}(t)$ (\ref{eq:Pc}), which behaves as $\mathcal{P}_{\mathrm{c}}(t) \approx 1-\omega\sqrt{Dt}$ at short times and $\mathcal{P}_{\mathrm{c}}(t) \approx c_{1} e^{-Dt/\lambda_{1}}$ at long times \cite{ignacio_2017,grebenkov_2020}, where $\omega$, $c_{1}$ and $\lambda_{1}$ are positive constants with $c_{1}\neq 1$ and $\lambda_{1}\neq\lambda$ as is the case for the exponential model (\ref{eq:exponential_model}).} 

Potential avenues for future work include accounting for a non-uniform initial distribution of particles, exploring other functional forms for approximating $\mathcal{P}(t)$ parameterised by one or more parameters and considering additional mechanisms such as drift or decay in the diffusive transport system. 

\appendix
\renewcommand{\theequation}{A.\arabic{equation}}
\setcounter{equation}{0}
\section{Averaging the diffusion equation}
\label{sec:averaging}
Obtaining a differential equation satisfied by $\mathcal{P}_{\mathrm{c}}(t)$ by averaging the continuum model (\ref{eq:cm_pde})--(\ref{eq:cm_bc2}) is only possible for the special case of reflecting boundary conditions at $x = \ell_{0}$ and $x=\ell_{1}$. To see why,
applying the averaging operator (\ref{eq:Pc}) to the diffusion equation (\ref{eq:cm_pde}) yields:
\begin{align}
\label{eq:cavg_de}
\frac{\text{d}\mathcal{P}_{c}}{\text{d}t} &= \frac{dD}{\ell_{1}^{d}-\ell_{0}^{d}}\left[\ell_{1}^{d-1}\frac{\partial c}{\partial x}(\ell_{1},t) - \ell_{0}^{d-1}\frac{\partial c}{\partial x}(\ell_{0},t)\right]\!.
\end{align}
For the trivial case where both boundary conditions (\ref{eq:cm_bc1})--(\ref{eq:cm_bc2}) are reflecting ($a_{0} = a_{1} = 0$ and $b_{0} = b_{1} = 1$), the differential equation (\ref{eq:cavg_de}) simplifies to $\text{d}\mathcal{P}_{c}/\text{d}t = 0$ and hence $\mathcal{P}_{\mathrm{c}}(t) = 1$ for all time as expected as no particles exit the system. This strategy fails, however, if either or both of the boundary conditions are absorbing or semi-absorbing ($a_{0}=1$ and/or $a_{1}=1$), as it is not possible to eliminate the dependence of equation (\ref{eq:cavg_de}) on $c(x,t)$ or its spatial derivative.

\renewcommand{\theequation}{B.\arabic{equation}}
\setcounter{equation}{0}
\section{Integral expressions}
\label{sec:integrals}
The exponential and Weibull models for the annulus and spherical-shell systems, outlined in Sections \ref{sec:exponential_annulus} and \ref{sec:weibull_annulus} respectively, depend on various geometrical constants, $I_{1}$, $I_{2}$, $I_{3}$ and $I_{4}$, which are expressed in terms of definite integrals for succinctness. Closed-form expressions for these integrals depend on the dimension $d$ and are given below:
\begin{align*}
I_{1} &= \begin{cases}\displaystyle  \ell_{1}-\ell_{0}, & \text{if $d = 1$},\\
\displaystyle \log_{e}(\ell_{1}/\ell_{0}), & \text{if $d = 2$},\\
\displaystyle \frac{\ell_{1}-\ell_{0}}{\ell_{0}\ell_{1}}, & \text{if $d = 3$},
\end{cases}\\
I_{2} &= \begin{cases}\displaystyle  \frac{(\ell_{1}-\ell_{0})^{2}}{2}, & \text{if $d = 1$},\\
\displaystyle \frac{2\ell_{1}^{2}\log_{e}(\ell_{1}/\ell_{0}) - (\ell_{1}^{2}-\ell_{0}^{2})}{4}, & \text{if $d = 2$},\\
\displaystyle \frac{(\ell_{1}-\ell_{0})^{2}(\ell_{0}+2\ell_{1})}{6\ell_{0}}, & \text{if $d = 3$},
\end{cases}\\
I_{3} &= \begin{cases}\displaystyle  \frac{(\ell_{1}-\ell_{0})^{3}}{6}, & \text{if $d = 1$},\\
\displaystyle \frac{(\ell_{0}^{2}+\ell_{1}^{2})\log_{e}(\ell_{1}/\ell_{0}) - (\ell_{1}^{2}-\ell_{0}^{2})}{4}, & \text{if $d = 2$},\\
\displaystyle \frac{(\ell_{1}-\ell_{0})^{3}}{6\ell_{0}\ell_{1}}, & \text{if $d = 3$},
\end{cases}\\
I_{4} &= \begin{cases}\displaystyle  \frac{(\ell_{1}-\ell_{0})^{4}}{24}, & \text{if $d = 1$},\\
\displaystyle \frac{4\ell_{1}^{2}(\ell_{1}^{2} + 2\ell_{0}^{2})\log_{e}(\ell_{1}/\ell_{0}) + \ell_{0}^{4}-5\ell_{1}^{4}+4\ell_{0}^{2}\ell_{1}^{2}}{64}, & \text{if $d = 2$},\\
\displaystyle \frac{(\ell_{1}-\ell_{0})^{4}(\ell_{0}+4\ell_{1})}{120\ell_{0}}, & \text{if $d = 3$}.
\end{cases}
\end{align*}

\vspace*{-2ex}
\bibliography{references}
\bibliographystyle{unsrt}

\end{document}